\documentclass{svjour2}
\usepackage{graphics}
\begin{document}
\newcommand{\beq}{\begin{equation}}
\newcommand{\eeq}{\end{equation}}
\bibliographystyle{apsrev}

\title{Comment on ``On Visibility in the Afshar Two-Slit Experiment"}

\author{Tabish Qureshi}
\institute{Department of Physics, Jamia Millia Islamia\\
New Delhi-110025, India.\\
\email{tabish.ph@jmi.ac.in}}


\maketitle

\begin{abstract}

Recently Kastner has analyzed the issue of visibility in a modified
two-slit experiment carried out by Afshar et al, which has been a subject
of much debate. Kastner describes a thought experiment which is claimed to
show interference with hundred percent visibility and also an ``apparent"
which-slit information. We argue that this thought experiment does not
show interference at all, and is thus not applicable to the Afshar experiment.

\end{abstract}

An experiment which
claims to violate Bohr's complementarity principle, proposed and carried
out by Afshar et al \cite{afsharfp}, is a subject of current debate.
Basically, it
consists of a standard two-slit experiment, with a converging lens behind
the conventional screen for obtaining the interference pattern. Although
If the screen is removed, the light passes through the lens and produces
two images of the slits, which are captured on two detectors $D_A$ and
$D_B$ respectively. Opening only slit $A$ results in only detector $D_A$
clicking, and opening only slit $B$ leads to only $D_B$ clicking. Afshar
argues that the detectors $D_A$ and $D_B$ yield information about which
slit, $A$ or $B$, the particle initially passed through. If one places a
screen before the lens, the interference pattern is visible.

Conventionally, if one tries to observe the interference pattern, one
cannot get the which-way information. Afshar has a clever scheme for
establishing the existence of the interference pattern without actually
observing it.  First the exact location of the dark fringes are noted by
observing the interference pattern. Then, thin wires are placed in the
exact locations of the dark fringes. The argument is that if the
interference pattern exists, sliding in wires through the dark fringes will
not affect the intensity of light on the two detectors. If the interference
pattern is not there, some photons are bound to hit the wires, and get
scattered, thus reducing the photon count at the two detectors. This way,
the existence of the interference pattern can be established without
actually disturbing the photons in any way.  Afshar et al carried out the
experiment and found that sliding in wires in the expected locations of the
dark fringes, doesn't lead to any significant reduction of intensity at the
detectors. Hence they claim that they have demonstrated a violation of
complementarity.

Recently, Kastner has addressed the issue of interference visibility in
the Afshar experiment \cite{kastner09}. Kastner believes that the essence of
the Afshar experiment is captured by a thought experiment discussed by Srikanth 
\cite{srikanth} in the context of complementarity. Kastner analyzed this
two-slit experiment in which there is an
additional internal degree of freedom of the detector elements
(which can be considered a “vibrational” component). The
particle + detector state evolves from the slits to the final screen with
initial detector state $|0\rangle$.  The detector spatial basis states
$|\phi_x\rangle$ and vibrational basis states $|v_U\rangle$ and $|v_L\rangle$
(corresponding to the particle passing through the upper and lower slit,
respectively) are activated. This evolution, from the initial state to the
detected particle, is given by
\begin{equation}
{1\over \sqrt{2}}(|U\rangle+|L\rangle)|0\rangle
\rightarrow \sum_x |x\rangle \left\{a_x |\phi_x\rangle|v_U\rangle + 
b_x |\phi_x\rangle|v_L\rangle\right\},
\end{equation}
where amplitudes $a_x$ and $b_x$ depend on wave number, distance, and slit
of origin, and $|x\rangle$ are final particle basis states.
Upon detection at a particular location $x$, one term remains from the sum
on the right-hand side of (1):
\begin{equation}
 |x\rangle \left\{a_x |\phi_x\rangle|v_U\rangle + 
b_x |\phi_x\rangle|v_L\rangle\right\}.
\end{equation}

Kastner argues that the result of this experiment is even more dramatic
than that of the Afshar experiment, because visibility is hundred percent
since a fully articulated interference pattern has been irreversibly
recorded - not just indicated indirectly - and yet a measurement can
be performed later, that seems to reveal “which slit” the
photon went through.

However, this argument is not correct, as can be seen from the following.
Suppose there were no ``vibrational states", then the term which remains
from the sum in (1) would be given by
\begin{equation}
 |x\rangle \left\{a_x |\phi_x\rangle + b_x |\phi_x\rangle\right\}.
\end{equation}
The probability density of detecting the particle at position $x$ is then
given by
\begin{equation}
 P(x) = \left\{|a_x|^2 + |b_x|^2 + a_x^*b_x + a_xb_x^* \right\}
\langle\phi_x|\phi_x\rangle,
\end{equation}
where the last two terms in the curly brackets denote interference.

One the other hand, the probability density of detecting the particle at
position $x$, in the presence of ``vibrational states" is given by
\begin{eqnarray}
 P(x) &=& \{|a_x|^2\langle v_U|v_U\rangle + |b_x|^2\langle v_L|v_L\rangle 
+ a_x^*b_x\langle v_U|v_L\rangle + a_xb_x^*\langle v_L|v_U\rangle \}
\langle\phi_x|\phi_x\rangle \nonumber\\
&=& \left\{|a_x|^2 + |b_x|^2 \right\} \langle\phi_x|\phi_x\rangle,
\end{eqnarray}
where the interference terms are killed by the orthogonality of $|v_U\rangle$
and $|v_L\rangle$.

So, contrary to the claim in \cite{kastner09}, this experiment does not
show any interference, although the ``vibrational states" do provide which-way
information. This is in perfect agreement with Bohr's complementarity
principle. It can show interference if $|v_U\rangle$
and $|v_L\rangle$ are not strictly orthogonal. However, in that case one
cannot extract any which-way information.

In conlcusion, we have shown that the thought experiment, described by
Kastner, does not show interference at all. What the experiment does
show is that if there exists which-way information in the state, there
is no interference pattern on the screen, in agreement with Bohr's
complementarity principle.

\end{document}